\documentclass[prletter,showpacs,twocolumn,typeset,superscriptaddress]{revtex4}
\usepackage{amsfonts}
\usepackage{amsmath}
\usepackage{amssymb}
\usepackage{graphicx}
\usepackage{epsfig}
\usepackage{citesort}

\usepackage{dcolumn}
\usepackage{bm}

\begin{document}

\preprint{APS}

\title{Petal-shape probability areas: complete quantum state discrimination}
\author{Luis Roa}
\author{Carla Hermann-Avigliano}
\author{Roberto Salazar}
\author{A. B. Klimov}
\affiliation{Departamento de F\'{\i}sica, Universidad de
Guadalajara, Revoluci\'{o}n 1500, 44420 Guadalajara, Jalisco,
Mexico.}
\author{B. Burgos}
\author{A. Delgado}
\affiliation{Center for Quantum Optics and Quantum Information,
Departamento de F\'{\i}sica, Universidad de Concepci\'{o}n, Casilla
160-C, Concepci\'{o}n, Chile.}
\date{\today}

\begin{abstract}
We find the allowed complex numbers associated with the inner
product of $N$ equally separated pure quantum states. The allowed
areas on the \textit{unitary complex plane} have the form of petals.
A point inside the petal-shape represents a set of $N$ linearly
independent (LI) pure states, and a point on the edge of that area
represents a set of $N$ linearly dependent (LD) pure states. For
each one of those LI sets we study the complete discrimination of
its $N$ equi-separated states combining sequentially the two known
strategies: first the unambiguous identification protocol for LI
states, followed, if necessary, by the error-minimizing measurement
scheme for LD states. We find that the probabilities of success for
both unambiguous and ambiguous discrimination procedures depend on
both the module and the phase of the involved \textit{inner product
complex number}. We show that, with respect to the phase-parameter,
the maximal probability of discriminating unambiguously the $N$
non-orthogonal pure states holds just when there no longer be
probability of obtaining ambiguously information about the prepared
state by applying the second protocol if the first one was not
successful.
\end{abstract}

 \maketitle

\section{Introduction}

The discrimination among different nonorthogonal pure quantum states
becomes a fundamental issue in Quantum Information and Computation
Theory \cite{Nielsen,Bergou}. There are two schemes in order to
achieve the task of recognizing nonorthogonal states. The first
scheme is the \textit{unambiguous quantum states discrimination}
about which has been written many interesting works: some generic
\cite{Ivanovic,Dieks,Peres,Chefles1,Chefles2,RSR} and others
concerned with its applications \cite{Bennett,Roa,Delgado,Hsu}. That
scheme requires that, whenever an outcome is projected after the
measurement process, one can, with nonzero probability, infer the
prepared state without error, this is, unambiguously. This can be
performed at the expense of allowing for a nonzero probability of
inconclusive outcomes. An \textit{optimum unambiguous states
discrimination} protocol is performed when the probability of
success is maximum. Thus, nowadays we know that, for discriminating
with certainty any one amongst $N$ nonorthogonal and linearly
independent (LI) pure states, it will be required to map or to
represent those states onto two orthogonal subspaces which are
called the \textit{conclusive-subspace} and the
\textit{inconclusive-subspace}. In the first $N$-dimensional
conclusive-subspace, each possible state to be discriminated has no
null component on only one state belonging to an orthonormal basis
whereas, in the inconclusive-subspace each state has no null
component on one state of a set of $N$ linearly dependent (LD)
states. In the second scheme, called the \textit{error-minimizing
measurement} protocol, the recognition of non-orthogonal states
accepts errors in the outcome results at the expense of that
inconclusive outcomes are not allowed. The advantage of this scheme
with respect to the former is that it can be applied to ambiguously
discriminate LD states \cite{Helstrom,Clarke,Andersson}. In the
second scheme an optimum ambiguous measurement process minimizes the
probability of making a wrong guess about the prepared state.

In this article we study the \textit{complete quantum states
discrimination} scheme among $N$ equally separated pure quantum
states. This \textit{complete scheme} consists in the consecutive
application of both protocols described above. To be specific, first
is applied the \textit{unambiguous quantum states discrimination}
protocol and, if it is not successful, then the system is mapped
onto the \textit{inconclusive-subspace} and then the
\textit{error-minimizing measurement} protocol is applied. This
\textit{complete scheme} allows getting all the obtainable
information about the prepared state. Before addressing that problem
we consider the characterization of the allowed complex numbers
associated with the inner product of the $N$ equally-separated pure
quantum states. We find that the allowed areas on the
\textit{unitary complex plane} have the form of petals and that a
point inside a petal-shape represents a LI set of $N$ pure states,
and a point on the edge of a petal represents a LD set of $N$ pure
states.

\section{\textit{N} Equally Separated States} \label{sec1}

Let us assume that the separation between two normalized states is
their inner product \cite{Barnett2}, with its module going from $0$
to $1$ and its phase being between $0$ and $2\pi$. The $0$ value of
the module corresponds to two orthogonal states and the $1$ value
refers to equal or parallel states. Two different states are always
LI and this property can be characterized completely by the fact
that the separation-module is different from $1$. However, for more
than two states the fact that the separation-modules among them are
different from $1$ guarantees neither the LI nor the LD property.
Thus an interesting question arises in this regard: given a set
$\verb"A"_N(\alpha)$ which has $N$ normalized and equi-separated
states, i.e.
\begin{equation}
\verb"A"_N(\alpha)\doteq\{|\alpha_1\rangle,|\alpha_2\rangle,\ldots,|\alpha_N\rangle
\colon\hspace{0.02in}\langle\alpha_k|\alpha_{k^\prime}\rangle=\alpha,\hspace{0.035in}\forall\hspace{0.05in}
k>k^{\prime}\},\nonumber
\end{equation}
what are the allowed values of $\alpha$? For which of those values
the $\verb"A"_N(\alpha)$ is a set of LI or LD states?

We know that $\verb"A"_N(\alpha)$ is a LI set when the identity
\begin{equation}
\sum_{k=1}^NA_k|\alpha_k\rangle=\textbf{null} \label{LIeq}
\end{equation}
is satisfied if and only if the $N$ coefficients $A_k$'s are all
zero, otherwise it is a LD set. Projecting Eq. (\ref{LIeq}) onto
each $|\alpha_k\rangle$ state we obtain an $N$ times $N$ homogeneous
linear equations system, being the $A_k$'s the $N$ unknown
quantities. From that homogeneous linear equations system one finds
that the $\verb"A"_N(\alpha)$ is a LI set when the following $N$
times $N$ matrix has determinant different from zero:
\begin{subequations}
\begin{eqnarray}
\mathcal{D}&=&\det\left(
\begin{array}{ccccc}
1 & \alpha & \alpha & \cdots & \alpha  \\
\alpha^{\ast} & 1 & \alpha & \cdots & \alpha  \\
\alpha^{\ast} & \alpha^{\ast} & 1 & \cdots & \alpha \\
\vdots & \vdots & \vdots & \ddots & \vdots  \\
\alpha^{\ast} & \alpha^{\ast} & \alpha^{\ast} & \cdots & 1
\end{array}
\right)_{N\times N}\label{detA0}\\
&=&\left[1+\alpha^{\ast}\sum_{k=0}^{N-2}\left(\frac{1-\alpha^{\ast}}{1-\alpha}\right)^{k}\right]\left(\alpha-1\right)^{N-1},\label{detA1}\\
&=&\frac{\alpha(\alpha^*-1)^N-\alpha^*(\alpha-1)^N}{\alpha^*-\alpha}.\label{detA2}
\end{eqnarray}
\label{detA}
\end{subequations}
The $\alpha$-roots of $\mathcal{D}$ correspond to sets of LD states.
Solutions such as $|\alpha|<0$, $|\alpha|>1$, and  $|\alpha|=$
\textit{complex number} are not allowed.

First of all, we consider $\alpha=x$ real. It is easy to show from
Eq. (\ref{detA1}) that, in this case, the $N$ states
$\{|\alpha_k\rangle\}$, are LI if and only if
\begin{equation}
-\frac{1}{N-1}<x<1.\nonumber
\end{equation}
For $x=1$ the $N$ states are LD and all equals whereas, for
$x=-1/(N-1)$ the $N$ states are LD and form a symmetric structure in
a $(N-1)$-dimensional subspace. The range $-1\leq x<-1/(N-1)$ is
forbidden for $N$ equally-separated normalized states. For instance,
a structure of three equi-separated states becomes LD just for
$x=-1/2$ which means that for this value of $x$ they can only lie on
a $2$-dimensional plane on which they are \textit{separated} by an
angle of $2\pi/3$. That family of three states is the well known
trine set \cite{Clarke,Andersson}. Four equi-separated states become
LD just for $x=-1/3$, i.e., they collapse symmetrically at a
$3$-dimensional subspace to form a tetrahedron.

Secondly, we consider the case of purely imaginary, i.e.
$\alpha=iy$. From Eq. (\ref{detA2}) we find that $\verb"A"_N(iy)$ is
a LI set for
\begin{equation}
-\frac{\sin\frac{\pi}{2N}}{\cos\frac{\pi}{2N}}<y<\frac{\sin\frac{\pi}{2N}}{\cos\frac{\pi}{2N}}.\nonumber
\end{equation}
The extreme values, $\pm\sin(\pi/2N)/\cos(\pi/2N)$, correspond to
$N$ equi-separated LD states whereas the range
$\sin(\pi/2N)/\cos(\pi/2N)<|y|\leq1$ is not allowed for normalized
states.

\begin{figure}[th]
\includegraphics[width=0.34\textwidth]{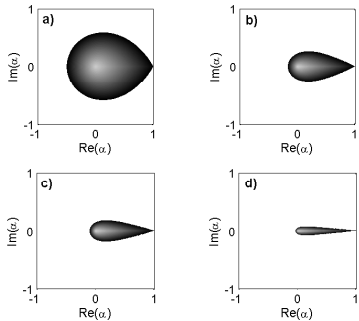}
\caption{Petal-shape areas corresponding to the allowed values of
$\alpha$ with the (\ref{Psucc}) probability in grey-degradation for:
a) $N=3$, b) $N=7$, c) $N=11$ and d) $N=31$. The black contour of
the grey-zone is the curve given by Eq. (\ref{absalpha2}), and white
color means the forbidden values.} \label{figure1}
\end{figure}

In the general case we find that $N$ equally-separated and
normalized LI states can only have inner product
$\alpha=|\alpha|e^{i\theta}$ which satisfies the constraint
\begin{equation}
0\leq|\alpha|<|\langle\alpha^{LD}_k(\theta)|\alpha^{LD}_{k^\prime}(\theta)\rangle|,\hspace{0.37in}0\leq\theta<2\pi,
\label{absalpha1}
\end{equation}
with
\begin{equation}
|\langle\alpha^{LD}_k(\theta)|\alpha^{LD}_{k^\prime}(\theta)\rangle|
=\frac{\sin\frac{\pi-\theta}{N}}{\sin\left(\theta+\frac{\pi-\theta}{N}\right)},
\label{absalpha2}
\end{equation}
whereas, at the contour defined by (\ref{absalpha2}) the $N$
equally-separated states are LD. Therefore, the allowed values of
the inner product, $\alpha$, for $N$ equi-separated states are
inside the region defined by the Eqs. (\ref{absalpha1}) and
(\ref{absalpha2}). Thus, each point in that zone represents a LI set
$\verb"A"_N(\alpha)$ and each point on the outline (\ref{absalpha2})
represents a LD set of $N$ equi-separated states. In other words,
for a given phase, $\theta$, there are infinite LI sets
$\verb"A"_N(|\alpha|e^{i\theta})$ and only one LD set,
$\{|\alpha^{LD}_k(\theta)\rangle$, $k=1,2,\ldots N\}$ and all those
families (LI's and the LD with $\theta$ fixed) preserve the phase of
their inner product changing only the module. Figure \ref{figure1}
shows those allowed region-values of $\alpha$ with grey-degradation
representing the (\ref{Psucc}) probability of success for: a) $N=3$,
b) $N=7$, c) $N=11$ and d) $N=31$. The clearest grey at $(0,0)$
means $1$, black contour means $0$, and white means the forbidden
values. Thus, it is worth emphasizing that along a radius ($\theta$
fixed) the sets $\verb"A"_N(|\alpha|e^{i\theta})$ preserve the
phase, changing only the module of the involved inner product and
just in the maximum allowed value (\ref{absalpha2}) of $|\alpha|$
the $N$ states become LD lying symmetrically on a $(N-1)-$
dimensional subspace. We can notice that the allowed surface in the
complex unitary-circle decreases as $N$ increases, going from a disk
for $N=2$, passing thru forms like a petal for $N>2$ up to tends to
be much closer to the positive real axis for $N\ggg2$. Because of
that reason we call those areas $(N>2)$ \textit{probability
petal-shapes}.

\section{Complete quantum pure states discrimination}  \label{sec2}

In the previous section we have characterized the sets of $N$
equi-separated states. In this section we shall describe the
\textit{complete quantum pure states discrimination} scheme for $N$
equally-separated and LI pure states.

Let us begin by supposing that a quantum system of interest is
prepared with probability $p_k$ in the state $|\alpha_k\rangle$
which belongs to the LI set $\verb"A"_N(\alpha)$. The
\textit{complete quantum pure states discrimination} scheme is
performed as follows: by means of a joint unitary operation,
$\hat{U}$, an ancillary system, initially in a known normalized
state $|\kappa\rangle_a$, is coupled to the system of interest in
such a way that:
\begin{equation}
\hat{U}|\alpha_k\rangle|\kappa\rangle_a=\sqrt{1-|s|^2}|k\rangle|\perp\rangle_a
+s|\alpha_k^{LD}(\theta)\rangle|\vdash\rangle_a,\label{expsu}
\end{equation}
where the two ancillary normalized states, $|\perp\rangle_a$ and
$|\vdash\rangle_a$ are orthogonal. The set
$\{|k\rangle,k=1,2,\ldots,N\colon\langle
k|k^\prime\rangle=\delta_{k,k^\prime}\}$ defines an orthonormal
basis in the unambiguous-subspace of the system of interest. The
$\{|\alpha_k^{LD}\rangle\}$ are $N$ LD states lying on the
ambiguous-subspace of the system of interest. That LD set allows
preserving the phase of the inner products and it is unique as we
saw in Section \ref{sec1}. The $s$ probability amplitude helps to
preserve the module of the inner product and, due to the symmetry,
it does not depend on $k$.

Hence, by performing a measurement process on the auxiliary system
on the $\{|\perp\rangle_a,|\vdash\rangle_a\}$ basis, the system of
interest is mapped with probability $P_s=1-|s|^2$ onto the
unambiguous-subspace and with probability $P_f=|s|^2$ onto the
ambiguous-subspace. Therefore, after that measurement procedure we
have the possibility of obtaining unambiguously the information
about the prepared state or of obtaining inconclusive information
about the prepared state by implementing the so called \textit{LD
states recognition with minimal-error} protocol on the
$\{|\alpha_k^{LD}\rangle\}$ set \cite{Helstrom,Clarke,Andersson}.

\subsection{Unambiguous discrimination probability}  \label{subsec1}

\begin{figure}[th]
\includegraphics[width=0.34\textwidth]{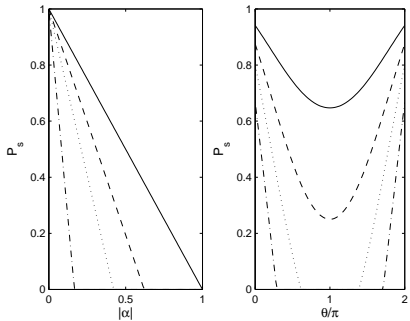}
\caption{Probability of success, $P_s$: a) as a function of
$|\alpha|$ for the different values of the phase $\theta=0$ (solid),
$\theta=\pi/11$ (dashes), $\theta=\pi/5$ (dots), and $\theta=\pi$
(dash-dots); b) as a function of $\theta$ for the different modules
$|\alpha|=1/17$ (solid), $|\alpha|=1/8$ (dashes), $|\alpha|=1/5$
(dots), and $|\alpha|=1/3$ (dash-dots). For these graphics we have
considered $N=7$.} \label{figure2}
\end{figure}

Since the inner product is preserved under a unitary transformation,
from Eq. (\ref{expsu}) we get:
\begin{eqnarray}
\langle\alpha_k|\alpha_{k^\prime}\rangle&=&|s|^2\langle\alpha_k^{LD}(\theta)|\alpha_{k^\prime}^{LD}(\theta)\rangle,\nonumber\\
|\alpha|e^{i\theta}&=&|s|^2|\langle\alpha_k^{LD}(\theta)|\alpha_{k^\prime}^{LD}(\theta)\rangle|e^{i\theta}\nonumber,
\end{eqnarray}
from where we obtain
\begin{equation}
|s|^2=\frac{|\alpha|}{|\langle\alpha_k^{LD}(\theta)|\alpha_{k^\prime}^{LD}(\theta)\rangle|}.
\end{equation}
Therefore, the probability of discriminating unambiguously the
$|\alpha_k\rangle$ state is
\begin{eqnarray}
P_k&=&p_k(1-|s|^2)\nonumber\\
&=&p_k\left(1-\frac{|\alpha|}{|\langle\alpha_k^{LD}(\theta)|\alpha_{k^\prime}^{LD}(\theta)\rangle|}\right),\nonumber
\end{eqnarray}
and the probability of discriminating unambiguously whichever be the
prepared state, becomes
\begin{eqnarray}
P_{s}&=&\sum_k P_k, \nonumber\\
&=&1-\frac{|\alpha|}{|\langle\alpha_k^{LD}(\theta)|\alpha_{k^\prime}^{LD}(\theta)\rangle|},\nonumber\\
&=&1-|\alpha|\frac{\sin\frac{\pi-\theta}{N}}{\sin\left(\theta+\frac{\pi-\theta}{N}\right)}.
\label{Psucc}
\end{eqnarray}
First of all we emphasize two limits: i) for $N=2$ and independently
of the $\theta$ value the (\ref{Psucc}) probability of success,
$P_s$, becomes the Peres's formula \cite{Peres}, $1-|\alpha|$, which
also holds for all $N\geq2$ when $\theta=0$, ii) the (\ref{Psucc})
probability of success is zero on the outline (\ref{absalpha2})
which is in agreement with the fact that a set of LD states can not
be unambiguously discriminated. We can also notice that the
(\ref{Psucc}) probability of success depends on both the phase and
the module of the involved $\alpha$. Specifically $P_s$ is linear
with respect to $|\alpha|$ having its maximum value, $1$, at
$|\alpha|=0$ and its minimum value, $0$, at the
$|\alpha|=|\langle\alpha_k^{LD}(\theta)|\alpha_{k^\prime}^{LD}(\theta)\rangle|$
cut-contour (see Fig. \ref{figure1}). Figure \ref{figure2}.a shows
that linear behavior considering $N=7$ for different phases:
$\theta=0$ (solid), $\theta=\pi/11$ (dashes), $\theta=\pi/5$ (dots),
and $\theta=\pi$ (dash-dots). On the other hand, we can notice that
for a given $N>2$: for $0\leq|\alpha|\leq1/(N-1)$, all the range
$0\leq\theta<2\pi$ is allowed, whereas for
$1/(N-1)<|\alpha|<|\langle\alpha_k^{LD}(\theta)|\alpha_{k^\prime}^{LD}(\theta)\rangle|$
the range of the phase is restricted having a wide window of values
non-allowed. Figure \ref{figure2}.b shows the (\ref{Psucc})
probability as a function of $\theta$ considering $N=7$ for
different modules of $\alpha$: $|\alpha|=1/17$ (solid),
$|\alpha|=1/8$ (dashes), $|\alpha|=1/5$ (dots), and $|\alpha|=1/3$
(dash-dots).

We emphasize that for a given and allowed module $|\alpha|$ the
maximal probability of discriminating conclusively $N(>2)$
equally-separated states holds for the phase $\theta=0$ and
curiously, in this case, the ambiguous-subspace is $1-$dimensional
since the $N$ equiseparated LD states,
$\{|\alpha^{LD}_k(\theta)\rangle\}$, are all equals. In other words,
in the optimal case all the possible information to be gotten about
the prepared states, is acquired unambiguously. On the other hand,
for $\theta\neq0$ each set, at the contour of the petal-shape, has
$N$ equiseparated LD states symmetrically distributed in a
($N-1$)-dimensional Hilbert subspace. In those cases
($\theta\neq0$), at the ambiguous-subspace there can even be
information about the prepared state and, even though this can no
longer be unambiguously obtained, it can be inconclusively obtained
by means of the so called \textit{LD states recognition with
minimal-error} protocol \cite{Helstrom,Clarke,Andersson}. Therefore,
we can say that the non-maximal cases ($\theta\neq0$) of obtaining
certainly the precedence of the state is at the expense of that part
of the information that can even be got with minimal-error.

\subsection{Ambiguous discrimination probability for equiseparated
LD states}  \label{subsec2}

When the unambiguous discrimination states process fails, the state
of the system of interest is projected, with probability $p_k$, onto
the state $|\alpha^{LD}_k(\theta)\rangle$, see Eq. (\ref{expsu}).
Those states belong to a LD set of $N$ equi-separated states
symmetrically distributed at a $(N-1)-$dimensional Hilbert subspace.
We would like to emphasize that for this analysis the $N$
equi-separated LD states must be different. In other words, our
analysis is valid for $\theta\neq0$ since the $\theta=0$ case is a
singularity because the $N$ states collapse to only one when
$\theta$ goes to $0$ (since
$|\langle\alpha_k^{LD}(0)|\alpha_{k^\prime}^{LD}(0)\rangle|=1$). In
addition, an obvious restriction to our analysis is $N>2$.

Therefore, in this \textit{minimal-error discrimination} scheme the
lowest error probability will be achieved by means of a measurement
process of the observable $\hat\Omega$ whose eigenstates
$\{|\omega_k\rangle,k=1,2,\ldots,N,\langle\omega_k|\omega_{k^\prime}\rangle=\delta_{k,k^\prime}\rangle\}$
are in a one-to-one correspondence with each possible state of the
LD set, $\{|\alpha_k^{LD}(\theta)\rangle,k=1,2,\ldots,N\}$, of the
system of interest. Let us suppose that each eigenstate
$|\omega_k\rangle$ has two components: one of them is lengthways
parallel to its corresponding $|\alpha_k\rangle$ and the other one
has a direction $|\Lambda_k\rangle$ which is orthogonal to the
\textit{ambiguous-subspace}, this is,
\begin{equation}
|\omega_k\rangle=\sqrt{1-|a|^2}|\Lambda_k\rangle+a|\alpha^{LD}_k(\theta)\rangle.\label{omega}
\end{equation}
We can notice that $|\alpha^{LD}_k(\theta)\rangle$ has component $a$
on its associated $|\omega_k\rangle$ and has component
$a|\langle\alpha^{LD}_k(\theta)|\alpha^{LD}_{k^\prime}(\theta)\rangle|$
over any $|\omega_{k^\prime}\rangle$ with $k>k^\prime$. From the
identity
\begin{eqnarray}
\sum_{m=1}^N\langle\alpha^{LD}_k(\theta)|\omega_m\rangle\langle\omega_m|\alpha^{LD}_k(\theta)\rangle&=&1\nonumber\\
|a|^2(N-1)|\langle\alpha^{LD}_k(\theta)|\alpha^{LD}_{k^\prime}(\theta)\rangle|^2+|a|^2&=&1,\nonumber
\end{eqnarray}
we obtain the square module of the $a$ probability amplitude:
\begin{equation}
|a|^2=\frac{1}{1+(N-1)|\langle\alpha^{LD}_k(\theta)|\alpha^{LD}_{k^\prime}(\theta)\rangle|^2}.\label{a}
\end{equation}
From Eqs. (\ref{omega}), (\ref{a}), and (\ref{absalpha2}) we can
infer that the probability, $P_{ci}$, that the state will be
correctly identified is:
\begin{eqnarray}
P_{ci}&=&\sum_{k=1}^Np_k|a|^2\nonumber\\
&=&\frac{1}{1+(N-1)\frac{\sin^2\frac{\pi-\theta}{N}}{\sin^2\left(\theta+\frac{\pi-\theta}{N}\right)}}.
\label{pci}
\end{eqnarray}
We notice that the minimum $P_{ci}$ probability holds for
$\theta\rightarrow0$ whereas the maximum value is reached at
$\theta=\pi$. In other words, $P_{ci}$ increases in the range
$0<\theta\leq\pi$ going from $1/N$ to $1-1/N$. On the other hand, in
the range $0<\theta\leq\pi$ and for $N\gg2$ the $P_{ci}$ probability
has the behavior
\begin{equation}
P_{ci}^{(N\gg2)}\approx1-\frac{(\theta-\pi)^2}{N\sin^2\theta}+O\left(\frac{1}{N^2}\right),
\nonumber
\end{equation}
From here we can see that the probability of correctly identifying
the $N$ equiseparated LD states comes closer to $1$ for $N\gg2$;
therefore, in this limit, the probability of erroneously inferring
the prepared state comes closer to $0$.

\begin{figure}[th]
\includegraphics[width=0.34\textwidth]{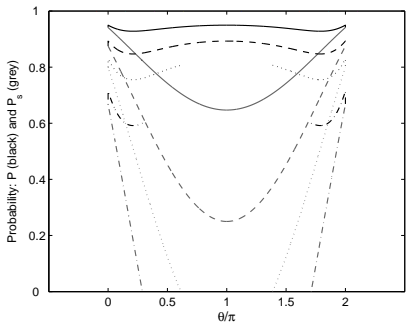}
\caption{The $P$ (black) and $P_s$ (grey) probabilities as functions
of $\theta$ for different values of $|\alpha|$: $|\alpha|=1/17$
(solid), $|\alpha|=1/8$ (dashed), $|\alpha|=1/5$ (dotted), and
$|\alpha|=1/3$ (dash-dotted). Here we have considered $N=7$.}
\label{figure3}
\end{figure}

According to the described \textit{complete quantum pure states
discrimination} scheme, see Eq. (\ref{expsu}), the \textit{total
probability}, $P$, \textit{of obtaining information} about the
prepared state of the system of interest becomes:
\begin{eqnarray}
P&=&P_s+(1-P_s)P_{ci}\nonumber\\
&=&1-\frac{|\alpha|(N-1)\frac{\sin\frac{\pi-\theta}{N}}{\sin\left(\theta+\frac{\pi-\theta}{N}\right)}}
{1+(N-1)\frac{\sin^2\frac{\pi-\theta}{N}}{\sin^2\left(\theta+\frac{\pi-\theta}{N}\right)}}.
\end{eqnarray}
We can define two more probabilities which give us knowledge about
the \textit{complete quantum pure states discrimination} scheme: The
$P_{err}=(1-P_s)(1-P{ci})$ which is the \textit{probability of
inferring the prepared state with error} and $P_{no-I}=1-P$ which
corresponds to the \textit{probability of not obtaining information}
about the prepared state of the system of interest. We notice that
the $P$ probability is linear and $P_{err}$, which is equal to
$P_{no-I}$, is quadratic with respect to the allowed $|\alpha|$. In
the Subsection \ref{subsec1} we find that for a given and allowed
$|\alpha|$ the maximum value of $P_s$ is reached at $\theta=0$ and
that probability decreases as $\theta$ goes from $0$ to $\pi$. On
the other hand, in Subsection \ref{subsec2} we see that $P_{ci}$
increases as $\theta$ goes from $0$ to $\pi$. However, since $P$ is
a composition of $P_s$ and $P_{ci}$ it is not easy to estimate its
behavior with respect to the $\theta$ parameter for a fixed
$|\alpha|$. Figure \ref{figure3} shows the $P$ (black) and $P_s$
(grey) probabilities as functions of $\theta$ for different values
of $|\alpha|$: $|\alpha|=1/17$ (solid), $|\alpha|=1/8$ (dashed),
$|\alpha|=1/5$ (dotted), and $|\alpha|=1/3 (dash-dotted)$. In order
to compare with results of Fig. \ref{figure2}.b we have considered
again $N=7$ for each curve of Fig. \ref{figure3}.

Figure \ref{figure3} shows the discontinuity of $P$ at $\theta=0$
and $\theta=2\pi$. We also notice that the \textit{total
probability}, $P$, \textit{of obtaining information} about the
prepared state is, in general, significatively higher than the
probability, $P_s$, of discriminating unambiguously the possible
prepared states. This also guarantees that the probability of
erroneously inferring the prepared state, $1-P$, becomes small.
Thus, the \textit{complete quantum pure states discrimination}
scheme allows obtaining all the possible information about state of
the system of interest.

\section{Conclusions} \label{sec3}
We have characterized the families of $N$ equally separated states
finding the allowed values of the involved inner product. We find
that the allowed surface in the complex unitary-circle decreases as
$N$ increases and it looks like petal-shapes. We studied the
unambiguous discrimination of those $N$ nonorthogonal and
equidistant pure quantum states, finding the probability of success.
That probability depends on both the module and the phase of the
inner product and curiously its maximum value, with respect to the
phase, arises just when the respective LD states become parallel in
such a way that ambiguous information can not be obtained. In all
the other cases the protocol can be complemented with a
\textit{minimal-error discrimination} scheme. In this form we have
proposed the \textit{complete quantum pure states discrimination}
scheme of $N$ equally-separated LI pure states which allows
obtaining all the possible information about the prepared state of
the system of interest.

\begin{acknowledgments}
Two of the authors (C.H.-A. and R.S.) thank Milenio ICM P06-067-F
for scholarship support. This work was supported by Grants: Milenio
ICM P06-067-F, FONDECyT N$^{\text{\underline{o}}}$ 1080535 and
1080383 and CONACyT N$^{\text{\underline{o}}}$45704.
\end{acknowledgments}

\end{document}